\documentclass[pra,twocolumn,english,showpacs,reprint,amssymb,aps, superscriptaddress, floatfix, 
longbibliography, nofootinbib]{revtex4-2}
\usepackage{graphicx,amsmath,amssymb}
\usepackage[colorlinks=true,allcolors=blue]{hyperref}
\usepackage{epstopdf}
\usepackage{bm}
\usepackage{upgreek}
\usepackage{babel}
\usepackage[normalem]{ulem}
\usepackage{orcidlink}
\usepackage{xcolor}

\newcommand{\tens}[1]{\bm{#1}}
\newcommand{\vect}[1]{\mathbf{#1}}

\newcommand{\vd}{\vect{d}}
\newcommand{\ve}{\vect{e}}

\newcommand{\vk}{\vect{k}}
\newcommand{\vp}{\vect{p}}

\renewcommand{\vr}{\vect{r}}

\newcommand{\vA}{\vect{A}}
\newcommand{\vD}{\vect{D}}

\newcommand{\vK}{\vect{K}}
\newcommand{\vF}{\vect{F}}
\newcommand{\vP}{\vect{P}}
\newcommand{\vR}{\vect{R}}

\newcommand{\cE}{\mathcal{E}}

\newcommand{\cH}{\mathcal{H}}
\newcommand{\cK}{\mathcal{K}}

\newcommand{\cW}{\mathcal{W}}
\newcommand{\cU}{\mathcal{U}}

\renewcommand{\Im}{\mathrm{Im}\,}

\begin{document}

\title{Bifurcation of laser-dressed band extrema in solids under intense low-frequency fields}

\author{A.~V. Flegel\,\orcidlink{0000-0003-1571-0745}}
\email{flegel@cs.vsu.ru}
\affiliation{Department of Physics, Voronezh State University, Voronezh 394018, Russia}

\author{Liang-Wen Pi\,\orcidlink{0000-0002-0430-1623}}
\email{lwpi@opt.ac.cn}

\affiliation{Laboratory of Ultrafast Optical Science and Technology, Xi'an Institute of Optics and Precision Mechanics, Chinese Academy of Sciences, Xi'an 710119, China}

\author{Liang-You Peng\,\orcidlink{0000-0002-3402-1577}}
\email{liangyou.peng@pku.edu.cn}
\affiliation{State Key Laboratory for Mesoscopic Physics and Frontiers Science Center for Nano-Optoelectronics, School of Physics, Peking University, Beijing 100871, China}

\author{M.~V. Frolov\,\orcidlink{0000-0001-6651-4619}}
\email{frolov@phys.vsu.ru}
\affiliation{Department of Physics, Voronezh State University, Voronezh 394018, Russia}
\affiliation{A.V. Gaponov-Grekhov Institute of Applied Physics, Russian Academy of Sciences, Nizhny Novgorod 603950, Russia}

\date{\today}

\begin{abstract}
We show that an intense low-frequency laser field can qualitatively modify the local extremal structure of an interband energy gap in a crystal. Using an instantaneous laser-dressed basis obtained from the nonperturbative diagonalization of the multiband crystal Hamiltonian, we derive a quartic normal form for the dressed gap near a high-symmetry point. At a critical field, the quadratic part of this normal form changes sign, and the gap experiences a bifurcation: a single symmetric minimum splits into two symmetry-related minima separated by a local maximum. We identify the microscopic origin of this bifurcation by decomposing the quadratic part into a homogeneous Stark contribution and a field-modified band-curvature contribution. The latter originates from the multiband redistribution of interband polarizations and provides the mechanism for the field-induced change of the gap extrema. A realization for the Kronig--Penney model demonstrates the same bifurcation at multiple high-symmetry points and for different interband gaps. As an illustration of its spectral consequences, we show that the bifurcation gives rise to an additional van Hove feature in the instantaneous interband susceptibility above the absorption edge. Our results establish a general mechanism by which strong-field dressing modifies the local geometry of the electronic spectrum.
\end{abstract}

\maketitle

\section{Introduction}

Strong electromagnetic fields provide a route to dynamically modifying the electronic structure of solids on ultrafast time scales. While the field-induced modification of optical absorption was established in the classical Franz--Keldysh picture~\cite{Franz1958, Keldysh1958}, the advent of intense few-cycle lasers has expanded the strong-field control of solids into a broad research field~\cite{Bao2022}. Of particular interest is the low-frequency regime, accessible with mid-infrared and terahertz sources, where the carrier frequency is small compared to characteristic interband energies. In this regime, the field varies slowly enough that the field-dressed electronic structure can be described quasi-statically. The validity of adiabatic state following, however, depends additionally on the relevant energy gaps and nonadiabatic couplings. The present work focuses on the instantaneous spectral structure itself.

Theoretically, the response of a crystal to a strong field has been described within several complementary frameworks. The Houston-state representation~\cite{Houston1940, Krieger1986}, based on the acceleration theorem, naturally describes intraband acceleration but leaves the instantaneous interband coupling explicit and nonperturbative in the evolution equations. The Floquet theory~\cite{Bukov2015, OkaKitamura2019} is powerful for periodic driving but is fundamentally unsuited for ultrashort, phase-stable pulses where the envelope varies on the optical-cycle scale. Fully numerical approaches based on the time-dependent Schr\"odinger equation, semiconductor Bloch equations, or the time-dependent density functional theory yield accurate \textit{ab initio} results but offer limited analytical insight into the underlying band geometry. To overcome these limitations, we extend the adiabatic approach developed in atomic strong-field physics~\cite{OTMPRA12,FlegPRA21,FrolovJPA2023} to spatially periodic potentials, thereby transferring the concept of instantaneous laser-dressed states to crystalline solids. This leads to an instantaneous laser-dressed basis that diagonalizes the multiband crystal Hamiltonian at the instantaneous crystal momentum and field strength, incorporating the nonperturbative field-induced interband mixing directly into the dressed band energies and matrix elements. Unlike Houston states, which diagonalize only the intraband motion, the dressed basis accounts for interband mixing in the electronic structure.

Experimentally, laser-dressed band structures have been directly observed via time-resolved angle-resolved photoemission, revealing Floquet--Bloch states and light-induced gaps on the surface of topological insulators~\cite{Wang2013, Mahmood2016} and, most recently, the light-driven band inversion in SnTe~\cite{Chassot2026}. In the low-frequency regime, the transient optical spectroscopy serves as a natural probe: the dynamical Franz--Keldysh effect~\cite{Jauho1996, Novelli2013} demonstrates that the instantaneous interband spectrum follows the driving field, and laser-dressed crystals exhibit robust purely optical signatures~\cite{Tiwari2025, Tiwari2023}. Recent studies have also explored field-induced modifications of band dispersion, such as the shifting or merging of Dirac nodes via the \textit{intraband} nonlinearity~\cite{Neufeld2023}. In contrast, the present work identifies a distinct mechanism in which the strong \textit{interband} mixing modifies the local structure of the interband gap. While intraband effects can modify dispersions through coherent acceleration, the interband mixing changes the field-dressed eigenstates and thereby the momentum dependence of the dressed gap.

Despite this progress, a fundamental question remains largely unexplored: Can an intense low-frequency field change not only the magnitude of an interband gap, but also its local geometry in momentum space, including the number, position, and character of its extrema? This question is also relevant to the optical response of the driven crystal, because changes in the extremal structure of the dressed gap modify the distribution of interband transition energies and can therefore generate new spectral features. We show that the answer is affirmative: a single symmetric minimum can bifurcate into two symmetry-related minima, while the original minimum becomes a local maximum. This bifurcation is driven not by a uniform Stark renormalization of the gap, but by a field-induced modification of its curvature arising from the multiband dressing.

To answer this question, we construct the instantaneous laser-dressed basis for a multiband crystal and derive the local normal form of the dressed gap near a high-symmetry point. We show that the field dependence of the local gap geometry is governed by quadratic and quartic gap tensors, whose evolution with the field strength determines the extremal structure. We establish that the field drives the quadratic part of the normal form through a sign change at a critical field, at which the single gap extremum bifurcates into two symmetry-related extrema. The microscopic origin of the sign change is identified through an exact decomposition into a homogeneous Stark contribution and a curvature contribution governed by the field-induced modification of the dressed-band curvatures. In the model considered here, the latter provides the dominant contribution and drives the bifurcation. We illustrate this mechanism numerically within the analytically tractable Kronig--Penney model, which provides a transparent setting for resolving the field-induced reshaping of the dressed-gap geometry. Although the present work uses a crystalline solid as the primary physical realization, the developed formalism relies solely on the discrete translational symmetry of the Hamiltonian and can therefore be applied more broadly to spatially periodic systems driven by strong low-frequency fields.

As a spectroscopic consequence of the bifurcation, the redistribution of the gap extrema modifies the interband susceptibility and produces an additional van Hove feature~\cite{VanHove1953} above the absorption edge. This provides an illustrative example of how the field-induced change in instantaneous gap geometry can enter an optical observable without requiring a dynamical treatment of interband transitions. The present work focuses on the instantaneous field-dressed electronic structure and establishes the field-controlled bifurcation of the local extremal structure of an interband gap together with its microscopic origin in the multiband dressing. This instantaneous spectral picture can subsequently be extended to the study of nonadiabatic interband dynamics in strong low-frequency fields.

Throughout the paper, atomic units are used unless stated otherwise.


\section{Adiabatic description of electron dynamics}
\subsection{Electron dynamics in the Houston representation}
\label{Sec:houston-basis}

We consider the dynamics of a single electron in a periodic structure driven by an intense low-frequency laser field. Throughout this work we employ the single-active-electron and dipole approximations and use the velocity gauge, which preserves the translational symmetry of the total Hamiltonian. 

The wave function $\Psi_{\vk}(\vr,t)$ of an electron with crystal momentum $\vk$ in the IR field satisfies the time-dependent Schr\"{o}dinger equation (TDSE):
\begin{align}
\label{TDSE}
& i\frac{\partial \Psi_\vk(\vr,t)}{\partial t}= \hat{H}\Psi_\vk(\vr,t),\\
\label{Hamiltonian:IR}
& \hat{H}=\frac{1}{2}\hat{\vP}_t^2+U(\vr),\quad \hat{\vP}_t=-i\nabla_\vr+\vA(t),
\end{align}
where $U(\vr)=U(\vr+\vR)$ is the spatially periodic (crystal) potential (for all $\vR$ from the Bravais lattice)  and $\vA(t)$ is the vector potential of the laser field.

The solution of Eq.~\eqref{TDSE} for a given $\vk$ can be obtained within the formalism of accelerated Bloch states or Houston functions $\varphi_{\vk,n}(\vr,t)$~\cite{Krieger1986} and presented as their superposition:   
\begin{equation}
\label{Psi-Bloch}
\Psi_\vk(\vr,t) = \sum_{n}a_{\vk,n}(t)\varphi_{\vk,n}(\vr,t).
\end{equation}
The wavefunction $\varphi_{\vk,n}(\vr,t)$ of the $n$-th band is the eigenstate of the instantaneous Hamiltonian $\hat{H}$:
\begin{equation}
\label{Houston-Eq}
\hat{H}\varphi_{\vk,n}(\vr,t) = E_n(\vK_t)\varphi_{\vk,n}(\vr,t),
\end{equation}
and can be expressed in terms of the field-free state $\psi_{\vk,n}(\vr)$ with energy $E_n(\vk)$ taken at the instantaneous field-dressed crystal momentum $\vK_t= \vk+\vA(t)$ [i.e., replacing $\vk\to\vK_t$]:
\begin{equation}
\label{Houston-psi}
\varphi_{\vk,n}(\vr,t) = e^{-i\vr\cdot\vA(t)}\psi_{\vK_t,n}(\vr) 
= e^{i\vk\cdot\vr}u_{\vK_t,n}(\vr),
\end{equation}
where $u_{\vK_t,n}(\vr) = u_{\vK_t,n}(\vr+\vR)$.  The time dependence of these states enters solely through the instantaneous crystal momentum $\vK_t$, which reflects the acceleration theorem for electrons in a periodic potential. The Houston functions therefore provide an exact description of intraband acceleration through the accelerated crystal momentum, but neglect the instantaneous interband coupling induced by the electric field.

Substituting Eq.~\eqref{Psi-Bloch} into the TDSE yields the exact evolution equations for the time-dependent expansion coefficients $a_{\vk,n}(t)$:
\begin{eqnarray}
\label{a(t)-Eq}
&&i\dot{a}_{\vk,n}(t) = E_n(\vK_t){a}_{\vk,n}(t) + V_{nn} {a}_{\vk,n}(t)\nonumber \\
&&+ \sum_{n'\ne n}V_{nn'}(t){a}_{\vk,n'}(t),
\end{eqnarray}
accompanied by the initial condition that the electron is in the valence band at $t \to -\infty$. Here $V_{nn'}(t)=-\langle\varphi_{\vk,n}|i\partial_t|\varphi_{\vk,n'}\rangle$ is the nonadiabatic coupling of adiabatic states $\varphi_{\vk,n}(\vr,t)$ and $\varphi_{\vk,n'}(\vr,t)$:
\begin{equation}
\label{Xnn'}
V_{nn'}(t) = \langle u_{\vK_t,n}|\hat{V}|u_{\vK_t,n'}\rangle 
= \vF(t)\cdot\vd_{nn'}(\vK_t),
\end{equation}
where $\hat{V} \equiv\hat{V}(t) = i\vF(t)\cdot\nabla_\vk$, $\vF(t)=- \dot\vA(t)$ is the laser field strength, and
\begin{equation}
\label{dnn'}
\vd_{nn'}(\vK_t) = \langle u_{\vK_t,n}|i\nabla_\vk| u_{\vK_t,n'} \rangle.
\end{equation}
The coupling matrix elements $V_{nn'}(t)$ characterize the laser-induced interaction between instantaneous adiabatic states. The off-diagonal components ($n \neq n'$) describe nonadiabatic interband transitions, whereas the diagonal components ($n = n'$) determine the Berry connection accumulated during the intraband motion. For crystals with inversion symmetry, $\vd_{nn}(\vk) = 0$, so the Berry connection vanishes identically. Hereafter we denote the instantaneous energy separation between bands by $\Delta_{nn'}(\vK_t) = E_n(\vK_t)-E_{n'}(\vK_t)$. The Houston representation diagonalizes the intraband motion exactly, while the entire nonadiabatic dynamics is governed by the instantaneous interband coupling matrix $V_{nn'}(t)$. In weak fields these couplings may be treated perturbatively. In the strong-field regime, however, they become comparable to the local band separation, and the perturbative description breaks down.

The central idea of the present work is to develop a new adiabatic basis that diagonalizes the multiband Hamiltonian at each instant. In the new basis, the nonperturbative laser-induced interband mixing is incorporated into the laser-modified electronic structure. Consequently, the description of electron dynamics proceeds through three successive representations: field-free Bloch states, accelerated Bloch (Houston) states, and instantaneous laser-dressed states. The first transformation (from Bloch to Houston states) exactly accounts for intraband acceleration, whereas the second (from Houston to laser-dressed states) accounts for the instantaneous interband mixing generated by the laser field.

\subsection{Laser-dressed adiabatic basis}
\label{Sec:adiabat-basis}

The Houston representation isolates the intraband acceleration but leaves the instantaneous interband coupling explicit in the evolution equations. In the low-frequency regime, the field-dressed Hamiltonian varies slowly compared with the characteristic interband energy scales. This allows the electronic structure to be described by an instantaneous eigenvalue problem, provided that the nonadiabatic coupling generated by the motion of the dressed basis remains small compared with the corresponding dressed-state energy separations. This leads naturally to a laser-dressed adiabatic basis, which forms the starting point for the analytical description of the dressed electronic structure developed below.

Our analytic approach is developed under the following assumptions. We consider a low-frequency laser field interacting with a periodic structure with band energies $E_n$ and energy separations $\Delta_{nn'}$ much larger than the laser-pulse carrier frequency,
\begin{equation}
\label{cond1}
|\Delta_{nn'}|\gg\omega.
\end{equation}
This condition establishes the low-frequency adiabatic regime in which interband transitions can be treated as nonadiabatic tunneling events in the instantaneous field~\cite{Zener34,Kane60}. 

In accordance with the key idea of the adiabatic (low-frequency) approach, we separate the fast oscillatory dynamics from the slow evolution of the amplitudes of $a_{\vk,n}(t)$. We factor out the dynamic phase by introducing $a_{\vk,n}(t)=\alpha_{nn_0}(t)\exp\!\left[-i\int^t\cE_{n_0}(\tau)\,d\tau\right]$, where the index $n_0$ labels the instantaneous dressed state and $\cE_{n_0}(t)$ denotes its instantaneous dressed energy, i.e., the time-dependent coefficients $\alpha_{nn_0}(t)$ form the adiabatic representation of the $n_0$-th laser-dressed state, $\upalpha_{n_0}\equiv\{\alpha_{nn_0}\}_n$. Substitution into Eq.~\eqref{a(t)-Eq} gives
\begin{equation}
\label{alpha_exact}
\sum_{n'}\cH_{nn'}(t)\,\alpha_{n'n_0} = \cE_{n_0}(t)\,\alpha_{nn_0} + i\dot{\alpha}_{nn_0},
\end{equation}
where the matrix $\cH_{nn'}(t)\equiv \cH_{nn'}(\vK_t,\vF_t)$ is the matrix form of the instantaneous effective Hamiltonian $\hat{\cH}(t)=\hat{H} - i\partial_t$ [where $\hat{H}$ is defined in Eq.~\eqref{Hamiltonian:IR}] in the Houston representation, including both the diagonal matrix elements of the Hamiltonian $\hat{H}$ and the contribution from the time dependence of the Houston basis:
\begin{equation}
\label{cHnn}
\cH_{nn'}(t) = E_n(\vK_t)\,\delta_{nn'} + \vF_t\cdot\vd_{nn'}(\vK_t),
\end{equation}
with $\vF_t=\vF(t)$.

The time derivative of the amplitudes $\alpha_{nn_0}$ (for all $n_0$) generates a nonadiabatic coupling between the dressed states:
\begin{equation}
\label{nonad_coupling}
\cW_{nn'} = i\upalpha_n^\dagger\dot{\upalpha}_{n'} = i\sum_{m}\alpha_{nm}^*\dot{\alpha}_{mn'}.
\end{equation}
Neglecting the nonadiabatic coupling leads to the standard formulation of the instantaneous eigenvalue problem:
\begin{equation}
\label{alpha:Eq}
\sum_{n'}\cH_{nn'}(t)\,\alpha_{n'n_0}(t) = \cE_{n_0}(t)\,\alpha_{nn_0}(t),
\end{equation}
which is controlled by the adiabaticity of the dressed-state following, i.e., the nonadiabatic coupling $\cW_{nn'}$ must be small compared with the dressed energy separations $\tilde{\Delta}_{nn'}=\cE_n-\cE_{n'}$:
\begin{equation}
\label{cond2}
|\tilde{\Delta}_{nn'}|\gg|\cW_{nn'}|.
\end{equation}
The condition~\eqref{cond2} for adiabatic following defines the adiabatic domain in which the instantaneous spectrum provides a controlled description of the driven electronic structure. The nonadiabatic coupling $\cW_{mn}$, which was neglected in the transition from the exact dynamical equation~\eqref{alpha_exact} to the algebraic eigenvalue problem~\eqref{alpha:Eq}, governs the evolution of $\upalpha_n(t)$. Its systematic inclusion constitutes a separate level of description that determines the range of applicability of the instantaneous spectral picture. 

In Eq.~\eqref{alpha:Eq}, all quantities depend on $t$ parametrically through the instantaneous crystal momentum $\vK_t$ and the field $\vF_t$. To simplify notations, we omit the time $t$ as an argument of the instantaneous band energies, adiabatic wave functions and matrix elements, if it is clear from the context, but keep the argument $\vK_t$ to indicate the specific value of the crystal momentum. Eq.~\eqref{alpha:Eq} can be solved within the stationary approach, i.e., by diagonalizing the Hamiltonian $\cH_{nn'}$. 

To specify the adiabatic conditions more precisely, we consider the time derivative of the amplitudes $\alpha_{nn_0}$, taking into account the parametric dependence on the crystal momentum $\vK_t$ and the field $\vF_t$:
\begin{equation}
\label{alpha_dot}
\dot{\alpha}_{nn_0} = -\vF_t\cdot\nabla_{\vk}\alpha_{nn_0} + \dot{\vF}_t\cdot\nabla_{\vF}\alpha_{nn_0}.
\end{equation}
The first term in Eq.~\eqref{alpha_dot} is the crystal-momentum gradient contribution associated with the motion of the dressed state in $\vk$-space. The second term describes the explicit variation of the dressed state with the instantaneous field strength. In the low-frequency limit, the explicit field-variation term carries an additional factor $\dot{\vF}\sim\omega F$. It can therefore be neglected when its associated nonadiabatic coupling remains small compared with the dressed-state energy separation,
\begin{equation}
\label{cond1-dressed}
|\tilde{\Delta}_{nn'}|\gg\omega|\vF_t\cdot\upalpha_{n}^\dagger\nabla_{\vF}\upalpha_{n'}|.
\end{equation}
Away from dressed-state near-degeneracies, this condition is naturally realized in the low-frequency limit. Thus, in the adiabatic domain within the low-frequency condition~\eqref{cond1-dressed}, the leading nonadiabatic coupling is the crystal-momentum gradient term, 
\begin{equation}
\label{Wnn}
\cW_{nn'}=-i\vF_t\upalpha_n^\dagger\nabla_{\vk}\upalpha_{n'}.
\end{equation}

The eigenvectors $\alpha_{nn_0}$ for all $n_0$ form the unitary transformation matrix, which determines the laser-dressed instantaneous basis $|\cU_{\vK_t,n_0}\rangle$:
\begin{equation}
\label{cU:exact}
|\cU_{\vK_t,n_0}\rangle = \sum_n \alpha_{nn_0}|u_{\vK_t,n}\rangle,
\end{equation}
and the dressed energy $\cE_{n_0}=\cE_{n_0}(\vK_t)$:
\begin{equation}
\label{cE:exact}
\cE_{n_0} = \sum_{n,n'} \alpha_{n_0n}^*\cH_{nn'}\alpha_{n'n_0}.
\end{equation}

Within the adiabatic approximation, the wave functions $\Phi_{\vk,n}(\vr,t)$,
\begin{equation}
\label{Phi:dressed}
\Phi_{\vk,n}(\vr,t) = e^{i\vk\cdot\vr}\,\cU_{\vK_t,n}(\vr,t),
\end{equation}
with energies $\cE_n$ form a set of eigenfunctions and eigenvalues of the instantaneous effective Hamiltonian $\hat{\cH}$: 
\begin{equation}
\label{eigenproblem:cV}
\hat{\cH}\Phi_{\vk,n}(\vr,t) = \cE_n(\vK_t)\Phi_{\vk,n}(\vr,t).
\end{equation}

This result allows us to construct the laser-dressed quantities, which will be used for further analysis. The dressed (instantaneous) polarization $\vD_{nn'}(\vK_t)$ on the basis of perturbed states $|\cU_{\vK_t,n}\rangle$ is defined by:
\begin{align}
\nonumber
\vD_{nn'}(\vK_t) &= \langle\Phi_{\vk,n}|i\nabla_\vk|\Phi_{\vk,n'}\rangle \\
&= \langle\cU_{\vK_t,n}|i\nabla_{\vk}|\cU_{\vK_t,n'}\rangle.
\label{Dnn}
\end{align} 
This quantity is related to the kinetic momentum matrix element $\vP(\vK_t)$, 
\begin{align*}
\vP_{nn'}(\vK_t) &= \langle\Phi_{\vk,n}|\hat{\vP}_t|\Phi_{\vk,n'}\rangle\\
&=\langle\cU_{\vK_t,n}|(-i\nabla_\vr + \vK_t)|\cU_{\vK_t,n'}\rangle,
\end{align*}
expressed in terms of the laser-dressed (instantaneous) energies $\cE_n(\vK_t)$ and the energy difference $\tilde{\Delta}_{nn'}(\vK_t)$:
\begin{equation}
\label{Pnn}
\vP_{nn'}(\vK_t) = 
\begin{cases}
\nabla_\vk\, \cE_n(\vK_t), & n=n',\\
i\tilde{\Delta}_{nn'}(\vK_t)\vD_{nn'}(\vK_t), & n\neq n'.
\end{cases}
\end{equation}
The relation~\eqref{Pnn} can be obtained taking into account the following equality: 
\[
\vP_{nn'}(\vK_t) = i\langle\Phi_{\vk,n}|[\hat{\cH},\vr]|\Phi_{\vk,n'}\rangle.
\]

Under the condition $|V_{nn'}| \ll |\Delta_{nn'}|$, the stationary perturbation theory (see, e.g.,~\cite{LL3}) can be used to obtain the expansion for $\alpha_{nn_0}(t)$: 
\begin{align}
\nonumber
\alpha_{nn_0} &= \delta_{nn_0} + (1-\delta_{nn_0})\biggl[\frac{\vF_t\cdot\vd_{nn_0}}{E_{n_0} - E_{n}} \\
&\quad + \sideset{}{'}\sum_{n'}\frac{(\vF_t\cdot\vd_{n n'})(\vF_t\cdot\vd_{n' n_0})}{(E_{n_0} - E_{n})(E_{n_0} - E_{n'})} + \cdots\biggr],
\label{series:alpha}
\end{align}
where the prime on the sum indicates that the singular term (with vanishing denominator) is excluded, and the arguments in $E_n= E_n(\vK_t)$ and $\vd_{nn'} = \vd_{nn'}(\vK_t)$ are omitted for simplicity. Using Eq.~\eqref{series:alpha}, we obtain the explicit expressions for $\cE_n$ to second order in $F_t$ and for $\vD_{nn'}$ to first order in $F_t$:
\begin{subequations}
\label{dressed:pt}
\begin{align}
\label{cE:dressed:2}
\cE_n &= E_n + \vF_t\cdot\vd_{nn} + \frac{1}{2}\vF_t\cdot\tens{\upchi}_{nn}\cdot\vF_t,\\
\label{D:dressed:1}
\vD_{nn'} &= \vd_{nn'} + \tens{\upchi}_{nn'}\cdot\vF_t.
\end{align}
\end{subequations}
Here $\tens{\upchi}_{nn'}$ is the linear susceptibility tensor, playing the role of a generalized polarizability tensor for the $n\to n'$ transition: 
\begin{equation}
\label{tens:chi}
\tens{\upchi}_{nn'} = \sideset{}{'}\sum_{m}\left[
\frac{\vd_{nm}\otimes \vd_{mn'}}{E_{n'}-E_m} 
+ \frac{\vd_{mn'}\otimes \vd_{nm} }{E_{n}-E_m}
\right].
\end{equation} 

The perturbation series in Eqs.~\eqref{series:alpha} and \eqref{dressed:pt} converges for $F d_{nn'} \ll \Delta_{nn'}$. This holds, for instance, for typical dielectrics with band gap $E_{\text{g}} \sim 5$--10~eV driven by mid-IR fields of a moderate intensity ($I \lesssim 10^{14}$~W/cm$^2$). For typical laser parameters ($I \lesssim 10^{13}$~W/cm$^2$), the laser-dressing effects are small and considered as perturbative corrections to the unperturbed Houston states. However, these effects require systematic analysis in the strong field regime.

The adiabatic condition~\eqref{cond2}, expressed through the gradient term in Eq.~\eqref{Wnn}, can be reformulated in explicit form, taking into account the perturbative expansion~\eqref{series:alpha}:
\begin{equation*}
\label{cond2:pert}
\left| \frac{\vF_t}{\Delta_{nn'}}\nabla_{\vk}\left(\frac{\vF_t\cdot\vd_{nn'}}{\Delta_{nn'}}\right) \right| \sim \left|\frac{V_{nn'}}{\Delta_{nn'}}\right|^2 \ll 1.
\end{equation*}
This inequality shows that the omitted gradient correction to the algebraic eigenvalue problem~\eqref{alpha:Eq} is of second order in the perturbative parameter $|V_{nn'}/\Delta_{nn'}|$ for the weak field regime.

The instantaneous eigenvalues obtained below define the laser-dressed electronic structure. Their local properties will be analyzed in the next section.


\section{Local geometry of laser-dressed bands}
\label{Sec:geometry}

\subsection{Local analytical representation of the dressed energy gap}
\label{Sec:kp-expansion}

We study the local geometry of the dressed electronic structure based on the proposed field‑dressed basis by considering an analytical representation of the dressed spectrum near the extremum of the instantaneous band separation. For definiteness, we focus on crystals with inversion symmetry, while the laser field is considered to be linearly polarized ($\ve$ is the unit polarization vector) along the principal crystal symmetry axis: 
\begin{equation*}
\vA(t)=\ve A(t),\quad \vF(t)=\ve F(t).
\end{equation*}
For a three-dimensional crystal, the laser field reduces the point group symmetry to the axial subgroup that leaves the axis invariant.

The general picture of the field-induced reshaping of the electronic band structure is illustrated in Fig.~\ref{fig:bandgap} for the Kronig--Penney model of a one-dimensional crystal (see Appendix~\ref{app:model} for details). Panels (a)--(c) show the laser-dressed bands $\cE_n(k)$ for three values of the instantaneous field strength $F_t=|\vF_t|$, while panel (d) relates these values to the temporal profile of the driving pulse. For the numerical illustrations, we parametrize the laser vector potential as
\begin{equation*}
A(t) = -\frac{F}{\omega} f(t)\sin\omega t, \quad 
f(t) = \sin^2\frac{\pi t}{\tau},
\end{equation*}
where $F$ is the peak amplitude of the field strength, $\omega$ is the carrier frequency, and $\tau$ is the pulse duration; the following values are used: $F =  1.52\times10^{-2}$~a.u.,  $\omega = 1.52\times10^{-2}$~a.u. (laser wavelength $\lambda = 3$~$\mu$m), and $\tau = 10\pi/\omega = 50$~fs.

Since the dressed basis diagonalizes the instantaneous effective Hamiltonian $\cH_{nn'}(K_t,F_t)$, the band structure at any instant within the pulse has the form displayed in Figs.~\ref{fig:bandgap}(a)--(c) with $F_t$ replaced by the instantaneous value $F(t)$; the markers $t_a$, $t_b$, $t_c$ indicate the instants at which $F(t)$ equals the field values used in the upper panels. With increasing $F_t$, the dressed bands are progressively reshaped: at $F_t=F^{(\rm cr)}$ the curvature of the dressed gap at the Brillouin-zone (BZ) boundary (the $X$ point) vanishes, and for $F_t>F^{(\rm cr)}$ the original extrema bifurcate into pairs of symmetry-related extrema --- the minimum of $\cE_1$ at $X$ splits into two minima separated by a newly formed local maximum, while the maximum of $\cE_0$ at $X$ splits into two maxima. An analogous bifurcation occurs at the zone center (the $\Gamma$ point), where the maximum of $\cE_1$ and the minimum of $\cE_2$ split similarly. The same behavior is seen in the higher bands. While illustrated here with the Kronig–Penney model, the mechanism of such a bifurcation is a generic property of the dressed band structure: it originates from the field-induced modification of the band curvatures and is not tied to any specific model or band pair. Whether the critical field $F^{(\rm cr)}$ is reached before material damage occurs, however, depends on the specific band structure. Panel (d) shows that $|F(t)|$ exceeds $F^{(\rm cr)}$ only in a finite time window around the pulse maximum; hence the bifurcated electronic structure is realized periodically, within each optical cycle. The corresponding evolution of the dressed gap $\tilde\Delta_{10}(k)=\cE_1(k)-\cE_0(k)$, whose single minimum at the BZ boundary flattens and eventually splits into two symmetry-related minima, is analyzed in the remainder of this section. The field at which this qualitative change occurs is identified below as the critical field $F^{(\rm cr)}$. 

\begin{figure*}[t]
\centering
\includegraphics[width=0.7\linewidth]{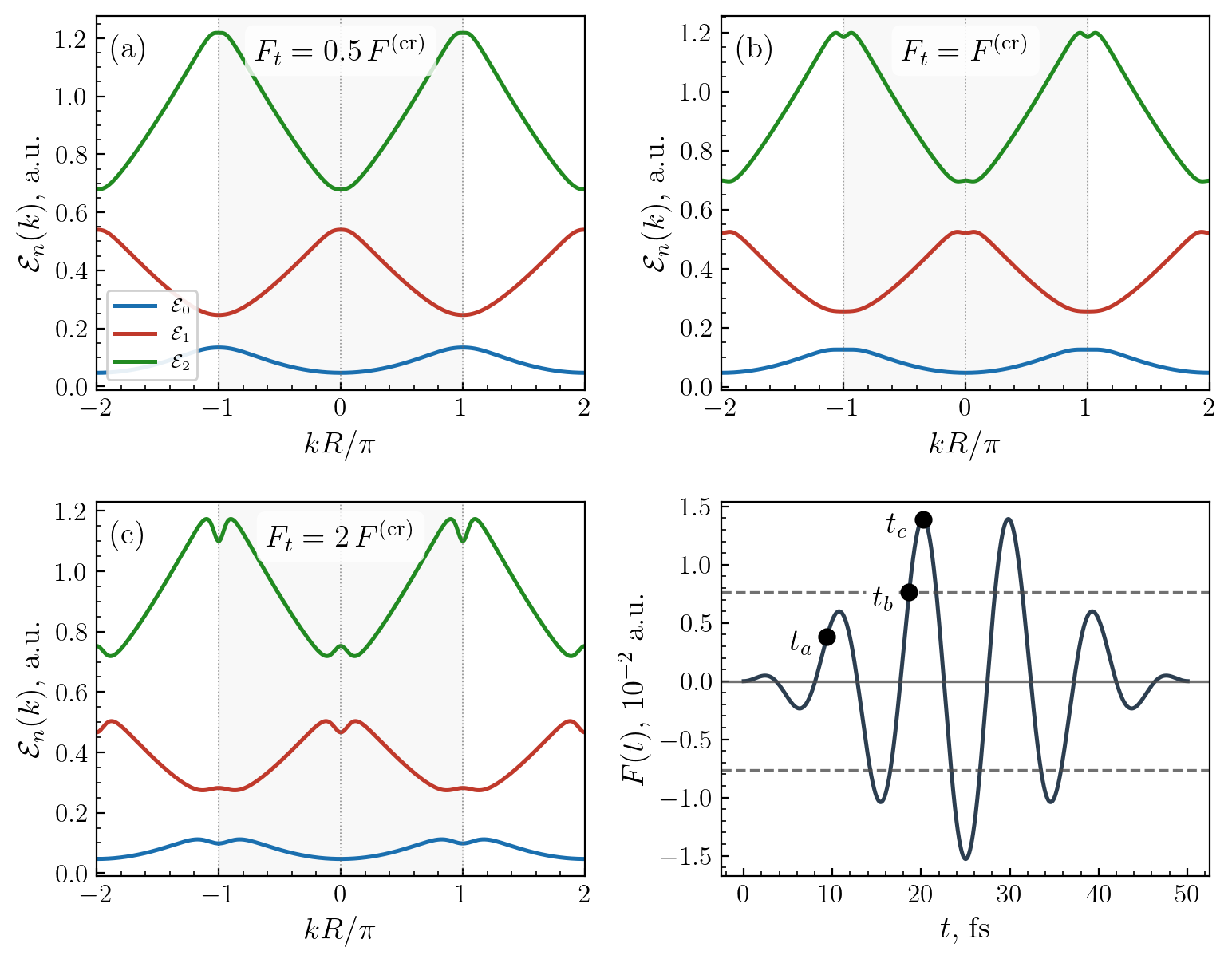}
\caption{(a)--(c) Laser-dressed bands $\cE_n(k)$, $n=0,1,2$, of the Kronig--Penney model for three values of the instantaneous laser field strength $F_t$ indicated in the panels; the shaded region marks the first Brillouin zone, and the vertical dotted lines indicate the high-symmetry points $\Gamma$ ($k=0$) and $X$ ($kR/\pi=\pm1$). With increasing $F_t$ the extrema of the dressed bands at $X$ and $\Gamma$ flatten and, for $F_t>F^{(\rm cr)}$, split into pairs of symmetry-related extrema. (d) Temporal profile of the driving pulse $F(t)$ of peak amplitude $2F^{(\rm cr)}$; the horizontal dashed lines mark $\pm F^{(\rm cr)}$, and the filled circles indicate the instants $t_a$, $t_b$, $t_c$ at which the instantaneous field equals the values used in panels (a)--(c). The model parameters correspond to a typical wide-bandgap crystal: lattice constant $R=6$~a.u., potential strength $g=0.4$~a.u.; the critical field is $F^{(\rm cr)}=7.6\times10^{-3}$~a.u. (corresponding to the intensity $I^{(\rm cr)}=2\times10^{12}$~W/cm$^2$).}
\label{fig:bandgap}
\end{figure*}

We first analyze the dressed gap in the vicinity of the original high-symmetry point $\vk_0$, which is determined by the equation:
\begin{equation}
\label{kp:k0-Eq}
\nabla_\vk\Delta_{10}(\vk)\Big|_{\vk=\vk_0} = 0.
\end{equation}
Since our analysis is applicable to any pair of neighboring bands, for simplicity we label the lower and upper bands as 0 and 1, respectively. We assume that $\vk_0$ lies on the principal crystal axis coinciding with the laser polarization direction (e.g., the $\Gamma$ or $X$ point for the [100] axis in a cubic crystal). Under this condition, the point $\vk_0$ remains a high-symmetry point of the dressed bands, and the dressed energies are even functions of the crystal-momentum shift 
\[
\delta\vk_0 = \vK_t-\vk_0.
\]

The instantaneous dressed states satisfy the eigenvalue problem in Eq.~\eqref{eigenproblem:cV} derived in Sec.~\ref{Sec:adiabat-basis}. Using this basis, we generalize the conventional $\vk\cdot\vp$ perturbation theory~\cite{Luttinger55, Kane57, Kane66} to the case of laser-dressed bands by applying a local expansion in the vicinity of a band-gap extremum. Let us denote the adiabatic wave function for a system with $N+1$ bands as $\Phi_\vk(\vr,t)$, and the sought eigenvalue as $\cE$. Substituting $\Phi_\vk(\vr,t)$ defined by the expansion,
\[
\Phi_\vk(\vr,t)=e^{i\vk\cdot\vr}\sum_{n=0}^{N}b_n\cU_{\vk_0,n}(\vr,t),
\]
into Eq.~\eqref{eigenproblem:cV}, we obtain the system of equations for the coefficients $b_n$:
\begin{subequations}
\begin{align}
\label{kp-system:N}
&\sum_{n'=0}^{N}\left(\cH_{nn'}-\cE\delta_{nn'}\right)b_{n'}=0,\\
&\cH_{nn'}=\left(\cE_n(\vk_0) +\frac{\delta\vk_0^2}{2}\right)
\delta_{nn'} + \vP_{nn'}(\vk_0)\cdot\delta\vk_0.
\label{kp-hamiltonian}
\end{align}
\end{subequations}
The terms in Eq.~(\ref{kp-hamiltonian}) have the following origin. (i)~The zeroth-order term $\cE_n(\vk_0)\delta_{nn'}$ is diagonal because the dressed basis diagonalizes the instantaneous effective Hamiltonian $\hat{\cH} = \hat{H} - i\partial_t$ at $\vk_0$; the bare coupling $V_{nn'}=\vF_t\cdot\vd_{nn'}(\vk_0)$ is incorporated into the dressed energies through this diagonalization and does not appear explicitly. (ii)~The linear diagonal term $\nabla_\vk\cE_n(\vk_0)\cdot\delta\vk_0$ vanishes at the high-symmetry points by the point-group symmetry of $\vk_0$. (iii)~The quadratic diagonal term $\delta\vk_0^2/2$ is the free-electron kinetic contribution, identical to the conventional $\vk\cdot\vp$ theory.

Within the partitioning method of L\"{o}wdin~\cite{Kane66, Lowdin1963} (see also Ref.~\cite{Feshbach1958}), we separate the two-band subsystem $A\equiv\{0,1\}$ with wave function\[
\Phi_\vk^{(A)}(\vr,t) = e^{i\vk\cdot\vr}\sum_{n\in A} a_n\cU_{\vk_0,n}(\vr,t),
\]
and take other bands ($n\notin A$) into account as perturbations. For materials with a wide bandgap, the effective two-band Hamiltonian $\cH_{nn'}^{(A)}$ (with $n,n'\in A$) is then given by:
\begin{equation}
\cH_{nn'}^{(A)} = \cH_{nn'} + \sum_{m\notin A}\frac{\cH_{nm}\cH_{mn'}}{\cE-\cH_{mm}}+\cdots.
\label{kp-hamiltonian:eff}
\end{equation}
We note that the effective Hamiltonian of subsystem $A$ becomes a function of $\cE$, which makes the eigenvalue problem nonlinear. For further analysis we evaluate the diagonal elements $\cH_{nn}^{(A)}$ at the unperturbed values $\cE\approx\cE_n(\vk_0)$, while the off-diagonal elements $\cH_{nn'}^{(A)}$ ($n\neq n'$) are evaluated at the midpoint energy $\cE\approx\overline{\cE}=[\cE_{0}(\vk_0)+\cE_{1}(\vk_0)]/2$, corresponding to the diabatic crossing of the two interacting levels. With determined matrix elements $\cH_{nn'}^{(A)}$, the eigenvalues $\cE$ for the two-band subsystem (i.e., $\cE_{0}$ and $\cE_{1}$) are found from the eigenvalue problem:
\begin{equation}
\label{kp-system:2}
\sum_{n'\in A}\left(\cH_{nn'}^{(A)} - \cE\delta_{nn'}\right) a_{n'} = 0.
\end{equation}
After diagonalization of $\cH_{nn'}^{(A)}$ in Eq.~\eqref{kp-system:2}, we obtain the expressions for $\cE_n$:
\begin{equation}
\label{kp:E}
\cE_n = \frac{1}{2}\left( \cH_{00}^{(A)} + \cH_{11}^{(A)} \pm  \tilde{\Delta}_{10} \right),
\end{equation}
where the sign $+$ $(-)$ relates to the band $n=1$ $(0)$, and the energy gap between neighboring dressed bands is:
\begin{align}
\label{kp:Delta-gen}
\tilde{\Delta}_{10} &= \sqrt{ \left( \Delta\cH \right)^2 + 4|\cH_{10}^{(A)}|^2}, \\
\nonumber
\Delta\cH &= \cH_{11}^{(A)}-\cH_{00}^{(A)}.
\end{align}

The corresponding dressed states $\cU_{\vK_t,n}$ of the considered two-band subsystem are:
\begin{subequations} 
\label{kp-states}
\begin{align}
\label{kp:Un0}
\cU_{\vK_t,0} &= -e^{-i\phi}\sin\frac{\theta}{2}\,\cU_{\vk_0,0} + \cos\frac{\theta}{2}\,\cU_{\vk_0,1},\\
\label{kp:Un0+1}
\cU_{\vK_t,1} &= \cos\frac{\theta}{2}\,\cU_{\vk_0,0} + e^{i\phi}\sin\frac{\theta}{2}\,\cU_{\vk_0,1},
\end{align}
\end{subequations}
where
\begin{equation}
\label{kp:theta-phi}
\phi = \arg\cH_{10}^{(A)},\quad \tan\theta = \frac{2|\cH_{10}^{(A)}|}{\Delta\cH}.
\end{equation} 

Using the analytic form~\eqref{kp-states} of the wave functions, we derive the analytic expression for the dressed dipole matrix element $\vD_{10}= \vD_{10}(\vK_t)$:
\begin{equation*}
\vD_{10} = - \frac{e^{-i\phi}}{2}\left(i\nabla_\vk\theta + \sin\theta\,\nabla_\vk\phi\right).
\end{equation*}
For a crystal with inversion symmetry, the geometric phase is quantized ($\phi=0$ or $\pi$) and its gradient vanishes identically. The interband dipole then reduces to:
\begin{equation}
\label{kp:D-general}
\vD_{10} =-i\frac{\Delta\cH\nabla_\vk\cH_{10}^{(A)} - \cH_{10}^{(A)}\nabla_\vk \Delta\cH}{\tilde{\Delta}^2_{10}}.
\end{equation}
The most important feature of the expression~\eqref{kp:D-general} is that the interband dipole $\vD_{10}(\vK_t)$ diverges at the complex branch points where the dressed gap vanishes, $\tilde{\Delta}_{10}(\vK_t)=0$. This divergence is a manifestation of the singularity of the adiabatic basis at band degeneracies, a well-known feature of the $\vk\cdot\vp$ theory~\cite{Luttinger55,Kane57,Kane66}. Importantly, the numerator in Eq.~\eqref{kp:D-general} is a smooth function of $\vK_t$ near the extremum $\vk_0$, while the entire singular behavior is contained in the denominator $\tilde{\Delta}^{-2}_{10}(\vK_t)$.

To clarify the analytical structure of the dressed energy difference $\tilde{\Delta}_{10}$ near the point $\vK_t = \vk_0$, first, we assume that in the limit $\vF_t\to 0$, there is a single minimum of $\Delta_{10}(\vk)$ at $\vk=\vk_0$ in the Brillouin zone (the same analysis is applicable to the general case of several well-separated extrema). Second, the laser-dressing effects can change the position of the gap minimum in the $\vk$-space leading to bifurcation of the minimum position for some value of the laser strength $F_t\geqslant F^{(\rm cr)}$. To specify the second point, we consider the vicinity of an isolated minimum $\vk_0$ of $\tilde{\Delta}_{10}(\vK_t)$ in the weak field regime $F_t<F^{(\rm cr)}$. 

Because the dressed basis diagonalizes the instantaneous effective Hamiltonian at $\vk_0$ and the crystal possesses inversion symmetry, the Taylor expansion of the exact dressed gap $\tilde{\Delta}_{10}(\vK_t)$ about $\vk_0$ contains only even powers of the crystal-momentum shift $\delta\vk_0$ and yields the local normal form (the same square-root structure follows from the two-band expansion of $\Delta\cH$ and $\cH_{10}^{(A)}$):
\begin{align}
\tilde{\Delta}_{10}(\vK_t) &= \tilde{\Delta}_{10}(\vk_0)
\bigl[ 1 + \delta\vk_0\cdot \tens{\upbeta}_{10}(\vk_0)\cdot\delta\vk_0 \nonumber \\
 &\quad + \delta\vk_0\delta\vk_0 : \tens{\upgamma}_{10}(\vk_0) : \delta\vk_0\delta\vk_0\bigr]^{1/2},
\label{kp:Delta2-tens}
\end{align}
where the single dot denotes standard tensor contraction, and the double dot denotes full contraction over all indices, i.e., 
\[
\delta\vk_0\delta\vk_0 : \tens{\upgamma}_{10} : \delta\vk_0\delta\vk_0 \equiv \sum_{i,j,l,m} \gamma_{10}^{ijlm} \delta k_i \delta k_j \delta k_l \delta k_m.
\] 

In Eq.~\eqref{kp:Delta2-tens}, $\tens{\upbeta}_{10}(\vk_0)$ is the quadratic gap:
\begin{align}
\tens{\upbeta}_{10} (\vk_0) &=\left.\frac{\nabla_{\vk}\otimes\nabla_{\vk}\tilde\Delta_{10}^2(\vK_t)}{2\tilde\Delta_{10}^2(\vk_0)}\right\vert_{\vK_t=\vk_0}\nonumber \\
&= \frac{\tens{\cK}_1(\vk_0)-\tens{\cK}_0(\vk_0)}{\tilde{\Delta}_{10}(\vk_0)},
\label{kp:tens-beta}
\end{align} 
where $\tens{\cK}_n$ is the dressed band-curvature tensor, i.e., the Hessian of the dressed energy (the inverse effective-mass tensor of the laser-dressed band):
\begin{equation}
\label{kp:curv-def}
\tens{\cK}_n(\vk)\equiv\nabla_{\vk}\otimes\nabla_{\vk}\cE_n(\vk),
\end{equation}
which reduces to the bare band curvature $\tens{\kappa}_n=\nabla_{\vk}\otimes\nabla_{\vk}E_n$ at $F_t\to0$. The curvature representation~\eqref{kp:tens-beta} directly relates $\tens{\upbeta}_{10}$ to the local geometry of the dressed gap and is most suitable for analyzing the evolution of the local gap profile with increasing laser field and for numerical calculations. An explicit form of the tensor $\tens{\cK}_n$ can be obtained directly from Eq.~\eqref{kp:E} using Eqs.~\eqref{kp-hamiltonian} and~\eqref{kp-hamiltonian:eff}:
\begin{align}
\label{kp:sumrule}
\tens{\cK}_n&=\tens{1}+2\sum_{m\neq n}\frac{\vP_{nm}\otimes\vP_{mn}}{\cE_n-\cE_m} \nonumber \\
&= \tens{1}+2\sum_{m\neq n}\tilde{\Delta}_{nm}\,\vD_{nm}\otimes\vD_{mn}.
\end{align}
Eq.~\eqref{kp:sumrule} leads to the microscopic representation for $\tens{\upbeta}_{10}$ and identifies the interband dipole matrix elements responsible for the same quantity. For the two-band approximation it gives the expression:
\begin{equation}
\label{kp:tens-beta-2band}
\tens{\upbeta}_{10} (\vk_0) = 4\,\vD_{10}(\vk_0)\otimes\vD_{01}(\vk_0),
\end{equation} 
which provides a simple physical interpretation of the weak-field behavior. However, being positive definite, this two-band estimate cannot by itself describe the sign change of $\tens{\upbeta}_{10}$. The sign change is driven by the contribution of higher bands ($m\geq 2$) to the sum-rule expression~\eqref{kp:sumrule}. Each higher band contributes the difference
\[ 
\tilde{\Delta}_{1m}\,\vD_{1m}\otimes \vD_{m1} - \tilde{\Delta}_{0m}\,\vD_{0m}\otimes \vD_{m0},
\] 
whose sign and magnitude are jointly governed by the energy ordering of the bands (through the signs of $\tilde{\Delta}_{1m}$ and
$\tilde{\Delta}_{0m}$, which is negative for two lower bands) and by the strength of the interband dipole matrix elements $\vD_{1m}$ and $\vD_{0m}$.

The fully symmetric quartic gap tensor $\tens{\upgamma}_{10}(\vk_0)$ in Eq.~\eqref{kp:Delta2-tens} originates from the higher order terms in the expansion of $\Delta\cH$ and $\cH_{10}^{(A)}$ in powers of $\delta\vk_0$ (the explicit form of $\tens{\upgamma}_{10}$ is not provided here). 

The local normal form of the dressed energy gap in the vicinity of the original high-symmetry point $\vk_0$ is given by Eq.~\eqref{kp:Delta2-tens}. This representation is sufficient as long as $\vk_0$ remains the unique local extremum. However, as the laser field increases, the evolution of this local geometry is entirely controlled by the field dependence of the tensors $\tens{\upbeta}_{10}$ and $\tens{\upgamma}_{10}$. For sufficiently strong fields $F_t>F^{(\rm cr)}$, new extrema may emerge. The location of these extrema is determined by equation 
\begin{equation}
\label{kp:tildek-Eq}
\nabla_\vk\tilde{\Delta}_{10}(\vK_t)\Big|_{\vK_t=\tilde{\vk}} =0, 
\end{equation}
or taking into account Eq.~\eqref{kp:Delta2-tens} by the following equation:
\begin{equation}
\label{kp:bifurcation-Eq}
\tens{\upbeta}_{10}(\vk_0) 
+ 2\tens{\upgamma}_{10}(\vk_0):(\tilde{\vk}-\vk_0)\otimes (\tilde{\vk}-\vk_0) = 0.    
\end{equation}

The structure of the gap tensors $\tens{\upbeta}_{10}$ and $\tens{\upgamma}_{10}$ is determined by the point-group symmetry of $\vk_0$. We consider the physically most relevant case in which the gap extremum $\vk_0$ lies on a principal crystal axis. This scenario describes a broad class of materials, including direct-gap semiconductors (where $\vk_0 = \Gamma$) and indirect-gap semiconductors with extrema on high-symmetry axes. For such symmetries, the small group of $\vk_0$ contains reflection planes perpendicular to the crystal axes, which ensure that both tensors are diagonal in the principal axes of the crystal (all off-diagonal components vanish). 

In the general case of a biaxial (orthorhombic) crystal (point group $D_{2h}$), the quadratic tensor has three independent diagonal components, while the quartic tensor $\tens{\upgamma}_{10}$ has up to nine independent components. If the principal axis containing $\vk_0$ has a rotational symmetry of order $N \geq 3$ (e.g., the [100] or [111] axis in a cubic crystal), the transverse plane becomes isotropic, and the tensors acquire the axial form. The quadratic tensor is characterized by two independent components:
\begin{equation}
\label{kp:beta-axial}
\tens{\upbeta}_{10} = \beta^\|_{10}\tens{P}^\| + \beta^\perp_{10}\tens{P}^\perp,
\end{equation}
where 
$\beta^\parallel_{10}$ is the longitudinal component, $\beta^\perp_{10}$ is the transverse component, and $\tens{P}^\|=\ve\otimes\ve$, $\tens{P}^\perp = \tens{1}-\ve\otimes\ve$. The quartic tensor has three independent components: $\gamma^\parallel_{10}$ (all four indices along $\ve$), $\gamma^{\perp\parallel}_{10}$ (two indices along $\ve$, two perpendicular), and $\gamma^\perp_{10}$ (all four indices perpendicular to $\ve$).

We focus on the case of longitudinal bifurcation, where new extrema emerge along the laser polarization direction. For the displacement $\tilde{\vk} - \vk_0 = (\tilde{k} - k_0)\,\ve$, Eq.~\eqref{kp:bifurcation-Eq} reduces to the scalar equation:
\begin{equation*}
\label{kp:bifurcation-long}
\beta^\parallel_{10} + 2\gamma^\parallel_{10} (\tilde{k} - k_0)^2 = 0,
\end{equation*}
which has two symmetry-related solutions:
\begin{equation}
\label{kp:bifurcation-x}
\tilde\vk_\pm = \vk_0\pm\sqrt{-\frac{\beta^\parallel_{10}}{2\gamma^\parallel_{10}}}\,\ve.
\end{equation}
The solutions exist when $\beta^\parallel_{10}$ and $\gamma^\parallel_{10}$ have opposite signs. The bifurcation occurs at the critical field $F^{(\rm cr)}$ defined by $\beta^\parallel_{10}(F^{(\rm cr)}) = 0$, at which point the original extremum at $\vk_0$ flattens and subsequently splits into the pair of symmetry-related extrema given by Eq.~\eqref{kp:bifurcation-x}, while the curvature of the gap at $\vk_0$ changes sign (original minimum becomes a local maximum and vice versa). 

The transverse stability of the newly formed minima requires that the second derivative of $\tilde{\Delta}_{10}^2$ with respect to any direction perpendicular to $\ve$ be positive at the new extrema. In the axial case, this condition reads:
\begin{equation}
\label{kp:stability-transverse}
\beta^\perp_{10} - 3\frac{\gamma^{\perp\parallel}_{10}\beta^\parallel_{10}}{\gamma^\parallel_{10}} > 0.
\end{equation}
If this condition is violated, the longitudinal minima become saddle points, and a transverse bifurcation scenario is realized, where new extrema emerge in directions perpendicular to the laser polarization. The analysis of transverse bifurcations requires consideration of the full tensor structure and is beyond the scope of the present work.

Once new extrema $\tilde{\vk}$ appear, it is convenient to introduce a new local coordinate measured from each field-induced extremum, 
\[\delta\tilde{\vk} = \vK_t - \tilde{\vk}.\]
In the vicinity of the new minima $\tilde{\vk}=\tilde{\vk}_+$ or $\tilde{\vk}=\tilde{\vk}_-$ for the near-bifurcation regime $F_t\gtrsim F^{(\rm cr)}$, the expansion of the squared dressed energy gap $\tilde{\Delta}^2_{10}(\vK_t)$ in powers of $\delta\tilde{\vk}$ to the second order gives:
\begin{equation}
\label{kp:Delta2-tens-new}
\tilde{\Delta}_{10}(\vK_t) = \tilde{\Delta}_{10}(\tilde{\vk}) \sqrt{ 1 + \delta\tilde{\vk}\cdot \tens{\upbeta}_{10}(\tilde{\vk})\cdot\delta \tilde{\vk} },
\end{equation}
where the quantities $\tilde{\Delta}_{10}(\tilde{\vk})$ and $\tens{\upbeta}_{10}(\tilde{\vk})$ can be evaluated either directly at the new minimum $\tilde{\vk}$ (both points $\tilde{\vk}_\pm$ give the same values due to symmetry with respect to the point $\vk_0$), or expressed in terms of their values at the original minimum~$\vk_0$ as: 
\begin{subequations}
\label{kp:tens-new}
\begin{align}
\label{kp:tens-new:Delta}
\tilde{\Delta}_{10}(\tilde{\vk}_\pm) &= C\tilde{\Delta}_{10}(\vk_0),\\        
\label{kp:tens-new:beta}
\tens{\upbeta}_{10}(\tilde{\vk}_\pm) &= \beta^\|_{10}(\tilde{\vk}_\pm)\tens{P}^\| + \beta^\perp_{10}(\tilde{\vk}_\pm)\tens{P}^\perp,\\
\label{kp:tens-new:beta-parallel}
\beta^\parallel_{10}(\tilde{\vk}_\pm) &= \frac{-2\beta^\parallel_{10}(\vk_0)}{C^2},\\
\label{kp:tens-new:beta-perp}
\beta^\perp_{10}(\tilde{\vk}_\pm) &= \frac{1}{C^2}\left(\beta^\perp_{10}(\vk_0) - 3\frac{\gamma^{\perp\parallel}_{10}(\vk_0)\beta^\parallel_{10}(\vk_0)}{\gamma^\parallel_{10}(\vk_0)}\right),\\
\label{kp:C}
C &= \sqrt{1 - \frac{\big[\beta^{\|}_{10}(\vk_0)\big]^2}{4\gamma^{\|}_{10}(\vk_0)}}.
\end{align}
\end{subequations}
Note that the numerator in Eq.~\eqref{kp:tens-new:beta-perp} coincides with the left-hand side of the transverse stability condition~\eqref{kp:stability-transverse}, so that the condition $\beta^\perp_{10}(\tilde{\vk}_\pm) > 0$ is equivalent to the requirement that the newly formed extrema be true minima rather than saddle points.

The dressed energy gap retains the same universal local structure in the vicinity of every instantaneous field-induced extremum [see Eq.~\eqref{kp:Delta2-tens-new}], while the corresponding quadratic gap tensor $\tens{\upbeta}_{10}(\tilde{\vk})$ and the position of the extremum $\tilde{\vk}$ become field dependent. The evolution of these quantities with increasing laser field determines the global restructuring of the dressed gap and is analyzed in the following subsection.

\begin{figure}[th]
\centering
\includegraphics[width=0.9\linewidth]{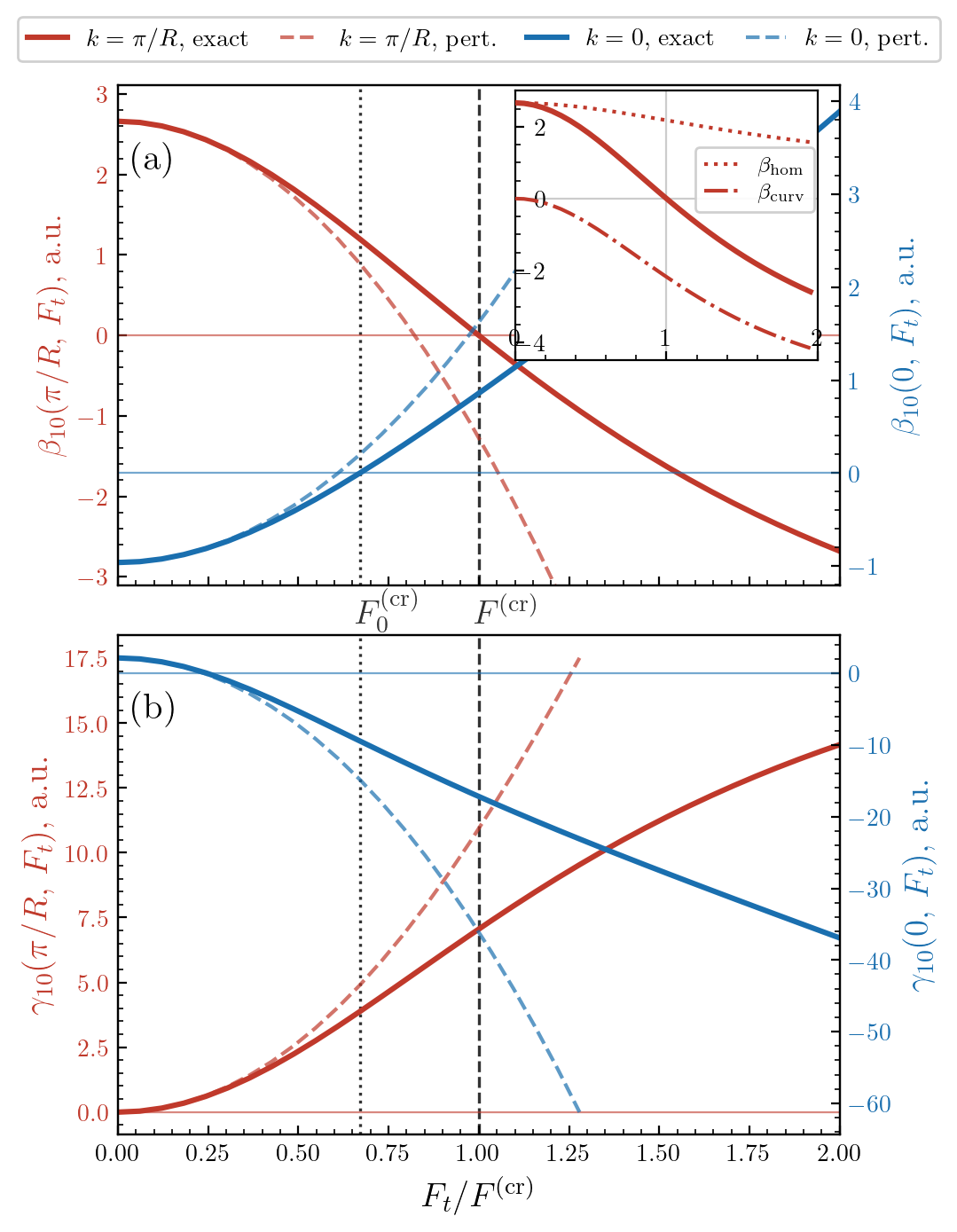}
\caption{(a) Quadratic coefficient $\beta_{10}(k_0,F_t)$ versus the instantaneous field strength $F_t$ for the high-symmetry points $k_0=\pi/R$ (red, left axis) and $k_0=0$ (blue, right axis). Solid lines: exact results from the curvature representation Eq.~\eqref{kp:tens-beta}; dashed lines: perturbative expansion Eq.~\eqref{tens-beta:pert}. Vertical dashed (dotted) line marks the critical field $F^{(\rm cr)}=7.6\times10^{-3}$~a.u. ($F^{(\rm cr)}_0=5.1\times10^{-3}$~a.u.) at which $\beta_{10}$ vanishes for $k_0=\pi/R$ ($k_0=0$). Inset: exact decomposition~\eqref{beta:representation} into the homogeneous $\beta_{\rm hom}$ (dotted) and curvature $\beta_{\rm curv}$ (dash-dotted) contributions, showing that the sign change of $\beta_{10}$ is driven by $\beta_{\rm curv}$. (b) Quartic coefficient $\gamma_{10}(k_0,F)$ for the same points; solid (dashed) lines denote exact (perturbative) results. Model parameters are the same as in Fig.~\ref{fig:bandgap}.}
\label{fig:beta-evolution}    
\end{figure}

\subsection{Laser field dependence of the dressed energy gap}
\label{Sec:gap-evolution}

The normal form of the laser-dressed energy gap in the vicinity of the high-symmetry point $\vk_0$, Eq.~\eqref{kp:Delta2-tens}, describes the local geometry of the energy gap. Its structure is dictated by the discrete translational and inversion symmetries of the crystal and does not depend on the specific band-structure model. The field dependence of the gap geometry in the vicinity of $\vk_0$ is determined by the quadratic ($\tens{\upbeta}_{10}$) and quartic ($\tens{\upgamma}_{10}$) gap tensors, whose variation with the field strength drives the bifurcation of the gap extremum when the longitudinal component $\beta_{10}^{\|}$ changes sign.

We illustrate the field dependence of the quadratic and quartic gap tensors using the Kronig-Penney model, for which both tensors reduce to the scalar coefficients $\beta_{10}(k,F_t)$ and $\gamma_{10}(k,F_t)$. Fig.~\ref{fig:beta-evolution}(a) shows the evolution of the quadratic coefficient $\beta_{10}(k_0,F_t)$ evaluated at the two high-symmetry points $k_0=0$ and $k_0=\pi/R$ with an increasing instantaneous field $F_t$. The curvature $d^2\tilde{\Delta}_{10}(k)/dk^2|_{k=k_0}=\cK_1(k_0)-\cK_0(k_0)$ at the $X$ and $\Gamma$ points of the BZ decreases monotonically and changes sign at the critical field $F=F^{(\rm cr)}$ for $k_0=\pi/R$ and $F_0^{(\rm cr)}$ for $k_0=0$, thereby explaining the appearance of the extrema bifurcation in Fig.~\ref{fig:bandgap}. The quartic coefficient $\gamma_{10}(k_0,F_t)$ [Fig.~\ref{fig:beta-evolution}(b)] also changes sign, but this occurs at fields well below the critical ones [e.g., $\gamma_{10}(\pi/R,F)$ is slightly negative at $F_t=0$ and turns positive at fields well below $F^{(\rm cr)}$]; consequently this weak-field sign change of $\gamma_{10}$ does not influence the bifurcation, which develops at $F_t\sim F^{(\rm cr)}$ where $\gamma_{10}$ is already a smooth positive-definite function of the field.

To identify the distinct contributions to the field-induced change of the quadratic gap tensor, we separate the change into a uniform rescaling and a curvature modification: the former relates to the uniform transformation of the energy gap with varying field strength (e.g., the Stark shift in the perturbative regime), while the latter corresponds to the geometrical reshaping of the bands. We represent the exact dressed gap as $\tilde{\Delta}_{10}(\vK_t) = \Delta_{10}(\vK_t) + \delta\Delta_{10}(\vK_t)$, where $\delta\Delta_{10}(\vK_t)$ is the field-induced gap modification. For an inversion-symmetric crystal both $\Delta_{10}(\vK_t)$ and $\tilde\Delta_{10}(\vK_t)$ have an extremum at $\vK_t=\vk_0$, and the Hessian is additive. Applying this decomposition for the curvature representation in Eq.~\eqref{kp:tens-beta} [see also \eqref{kp:curv-def} for the dressed curvature definition] and leaving the denominator in Eq.~\eqref{kp:tens-beta} unchanged, we obtain
\begin{subequations}
\label{beta:representation}
\begin{align}
\label{beta:repr:sum}
\tens\upbeta_{10} &= \tens\upbeta_{\rm hom} + \tens\upbeta_{\rm curv},\\
\label{beta:repr:hom}
\tens\upbeta_{\rm hom} &= \frac{\Delta_{10}(\vk_0)}{\tilde\Delta_{10}(\vk_0)}\tens\upbeta_{10}^{(0)},\\
\label{beta:repr:curv}
\tens\upbeta_{\rm curv} &= \frac{1}{\tilde\Delta_{10}(\vk_0)} \nabla_\vk \otimes \nabla_\vk \delta\Delta_{10}(\vK_t)\Big|_{\vK_t=\vk_0},
\end{align}
\end{subequations}
where $\tens\upbeta_{10}^{(0)}$ is the zero-field tensor at $\vK_t=\vk_0$. The two terms have distinct meanings. The homogeneous term~\eqref{beta:repr:hom} uniformly rescales $\tens\upbeta_{10}^{(0)}$ by the factor $\Delta_{10}/\tilde\Delta_{10}$, i.e., since the field increases the gap, $\tens\upbeta_{\rm hom}$ provides a uniform reduction of the zero-field nonparabolicity without altering its tensor structure. The curvature term~\eqref{beta:repr:curv} describes the genuine reshaping of the gap profile: it is the curvature of the field-induced modification $\delta\Delta_{10}$. Because $\delta\Delta_{10}(\vk)$ is maximal at $\vk_0$, where the interband coupling is strongest and the gap is smallest, the curvature $\nabla_\vk \otimes \nabla_\vk \delta\Delta_{10}$ is negative-definite. Physically, the field-induced gap modification $\delta\Delta_{10}$ forms a local ``hill'' in momentum space around $\vk_0$, so that its second derivatives are negative. The term $\tens\upbeta_{\rm curv}$ therefore reduces the gap curvature and, upon increasing the field, drives the eigenvalue $\beta_{10}^{\|}$ of $\tens\upbeta_{10}$ to zero, triggering the bifurcation.

In the weak-field regime the exact decomposition~\eqref{beta:representation} reduces to the perturbative form. Expanding $\tilde{\Delta}_{10} = \Delta_{10} + \delta\Delta_{10}^{(2)}+O(F^4)$ with the quadratic Stark shift [cf. Eqs.~\eqref{cE:dressed:2} and~\eqref{tens:chi}],
\[
\delta\Delta_{10}^{(2)} = \frac{1}{2}\vF_t\cdot(\tens\upchi_{11}-\tens\upchi_{00})\cdot\vF_t,
\]
and keeping terms to $O(F^2)$, the homogeneous term gives $\tens\upbeta_{\rm hom} = \tens\upbeta_{10}^{(0)} + \tens\upbeta_{\rm Stark}$, and the leading part of the curvature term is $\tens\upbeta_{\rm curv}^{(2)}$, so that
\begin{subequations}
\label{tens-beta:pert}
\begin{align}
\label{tens-beta:pert:sum}
\tens\upbeta_{10} &= \tens\upbeta_{10}^{(0)} + \tens\upbeta_{\rm Stark} + \tens\upbeta_{\rm curv}^{(2)},\\
\label{tens-beta:pert:stark}
\tens\upbeta_{\rm Stark} &= - \frac{\delta\Delta_{10}^{(2)}(\vk_0)}{\Delta_{10}(\vk_0)} \tens\upbeta_{10}^{(0)},\\
\label{tens-beta:pert:curv}
\tens\upbeta_{\rm curv}^{(2)} &= \frac{1}{\Delta_{10}} \nabla_\vk \otimes \nabla_\vk \delta\Delta_{10}^{(2)}(\vK_t)\Big|_{\vK_t=\vk_0}.
\end{align}
\end{subequations}
The perturbation expansion in Eqs.~\eqref{tens-beta:pert} explicitly demonstrates that the field dependence of $\tens{\upbeta}_{10}$ is controlled by the differential Stark shift of two neighboring bands ($\delta\Delta_{10}^{(2)}$): the homogeneous term $\tens\upbeta_{\rm Stark}$ gives a uniform reduction, whereas the curvature term $\tens\upbeta_{\rm curv}^{(2)}$ renormalizes the gap curvature through the field-induced modification of the diagonal susceptibilities, aggregating the field-induced modification of all interband couplings. 

In Fig.~\ref{fig:beta-evolution}(a) we compare the exact results based on Eq.~\eqref{kp:tens-beta} with the perturbative expansion Eq.~\eqref{tens-beta:pert}. The perturbative expression reproduces the exact $\beta_{10}(k_0,F_t)$ for $F_t\lesssim F^{(\rm cr)}$ and correctly predicts the sign change at the critical field, but progressively deviates at stronger fields, as expected for a second-order expansion. The inset separates the exact $\beta_{10}(\pi/R,F)$ into the homogeneous contribution $\beta_{\rm hom}$ given by Eq.~\eqref{beta:repr:hom}, which merely rescales the zero-field tensor, and the curvature contribution $\beta_{\rm curv}$ [Eq.~\eqref{beta:repr:curv}], which accounts for the field-induced modification of the gap curvature. For the model parameters considered here, the comparison shows that $\beta_{\rm curv}$ is the dominant mechanism driving $\beta_{10}$ to zero, providing the leading contribution to the field-induced reduction of $\beta_{10}$, while $\beta_{\rm hom}$ yields only a moderate uniform suppression. Fig.~\ref{fig:beta-evolution}(b) shows that the quartic coefficient $\gamma_{10}$ varies smoothly and monotonically across the bifurcation, confirming that the geometric modification of the gap is controlled by the quadratic coefficient.

\begin{figure}[t]
\centering
\includegraphics[width=\linewidth]{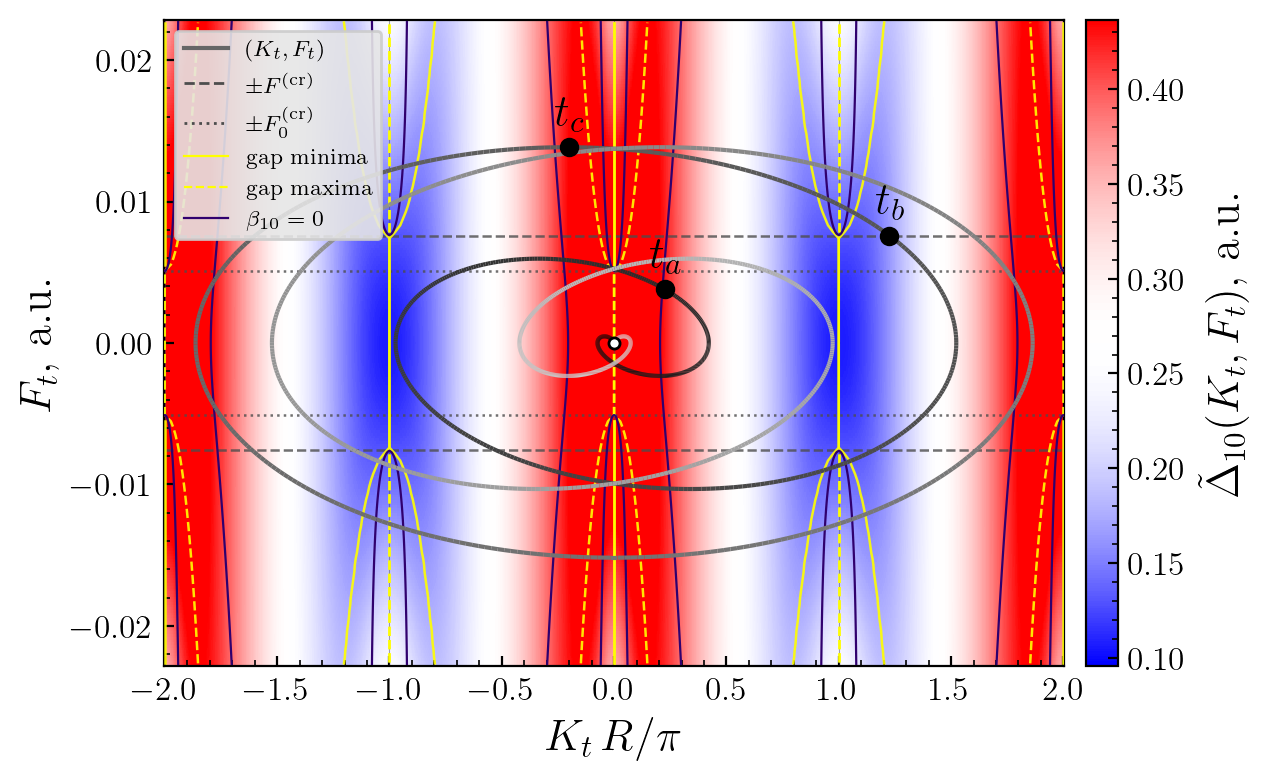}
\caption{Laser-dressed gap $\tilde\Delta_{10}(K_t,F_t)$ shown as a color-coded map. Yellow solid (dashed) curves trace the field-dependent positions of the local minima (maxima) of the gap; dark solid lines indicate where $\beta=\beta_{10}(K_t,F_t)=0$. Horizontal dashed and dotted lines indicate the critical fields $\pm F^{(\rm cr)}$ (for $k_0 = \pi/R$, the $X$ point---dashed line) and $\pm F^{(\rm cr)}_0$ (for $k_0 = 0$, the $\Gamma$ point---dotted line) at which the extrema bifurcate. The grayscale curve is the trajectory in the $(K_t,F_t)$ parameter plane of a driven electron (lighter points correspond to increasing $t$ with initial $K_t=k=0$); the filled circles mark the instants $t_a$, $t_b$, $t_c$ at which $F(t)$ equals the values used in Figs.~\ref{fig:bandgap} and \ref{fig:beta-evolution}. Model parameters as in Fig.~\ref{fig:bandgap}.}
\label{fig:gap_map}
\end{figure}

The dressed gap $\tilde\Delta_{10}(K_t,F_t)$ as a function of two key quantities $K_t$ and $F_t$ (i.e., the parameters of the field-dressed instantaneous effective Hamiltonian) is presented by a color-coded map in Fig.~\ref{fig:gap_map}. The dressed gap behaves similarly to the field-free case over the extended BZ for $|F_t|<F^{(\rm cr)}$, i.e., exhibits an alternating minimum-maximum pattern with a single minimum at $X$ and a single maximum at $\Gamma$. As $F_t$ reaches $F^{(\rm cr)}$ the extrema structure of the gap bifurcates by a pitchfork scenario (see yellow lines of extrema in Fig.~\ref{fig:gap_map}). The gap values at the extrema vary smoothly as the field crosses the critical values and remain non-zero over the whole $(K_t, F_t)$ plane. To illustrate the time-dependent dynamical reshaping of the energy gap as the field follows the laser pulse, we plot the parametric trajectory in the plane $(K_t,F_t)$ (the grayscale curve) for a driven electron with initial zero crystal momentum ($k=0$). As the field coordinate follows the laser pulse and the crystal momentum passes through the Brillouin zone, a trajectory goes through the sequence of instantaneous gap geometries. At $t_a$ ($F_{t_a}<F^{(\rm cr)}$) the electron moves in a gap with a single minimum at $X$, at $t_b$ ($F_{t_b}\simeq F^{(\rm cr)}$) it crosses the bifurcation point, and at $t_c$ ($F_{t_c}>F^{(\rm cr)}$) it explores the reshaped bifurcated geometry. Hence, the bifurcated structure is realized periodically within each optical cycle whenever $|F(t)|$ exceeds~$F^{(\rm cr)}$.

\subsection{Optical signature of the dressed-gap bifurcation}
\label{Sec:optical-signature}

The instantaneous laser-induced band reshaping may affect microscopic properties of a solid, such as the interband optical susceptibility. This quantity can be probed by a short UV pulse in the presence of an intense low-frequency pulse. In this two-color scheme, the short UV pulse serves as a probe field that selects a specific moment in the laser-induced adiabatic evolution of bands by varying the time delay. This scheme can be realized by superimposing an intense few-cycle terahertz pulse and a short UV pulse. As an example, we consider the instantaneous interband optical susceptibility probed at the time delay between two components, i.e., at $t=\tau$:
\begin{equation}
\label{chi-inst}
\tilde\chi(\Omega;F_\tau)=\int_{\rm BZ}\frac{dk}{2\pi}\,
\frac{|\tilde D_{10}(k,F_\tau)|^{2}}{\tilde\Delta_{10}(k,F_\tau)-\Omega-i\Gamma},
\end{equation}
where $\Gamma$ is a phenomenological broadening. Eq.~\eqref{chi-inst} is written in a one-dimensional form appropriate for the Kronig--Penney model. For a filled lower dressed band and an empty upper dressed band (the population factor is unity, consistent with a weak probe and the initial condition of an electron in the valence band), Eq.~\eqref{chi-inst} is the instantaneous (adiabatic) limit of the interband response, in which the crystal samples a fixed field $F_\tau$. Its imaginary part provides the absorptive interband response and, in the present model, is controlled by the extrema of the dressed gap weighted by the corresponding interband dipole.

According to Eq.~\eqref{chi-inst}, $\tilde\chi(\Omega;F_\tau)$ is given by a Brillouin-zone integral, so that its threshold structure directly reflects the extremal structure of $\tilde\Delta_{10}(k,F_\tau)$. Within the local one-dimensional normal form in Eq.~\eqref{kp:Delta2-tens}, a quadratic minimum $\tilde\Delta(k,F_\tau) - \tilde\Delta_0 \sim (k-k_0)^2$ [$\tilde\Delta_0 = \tilde\Delta(k_0,F_\tau)$] produces the standard one-dimensional edge singularity $\Im\tilde\chi\sim(\Omega-\tilde\Delta_0)^{-1/2}$, whereas at the critical field, where $\beta=0$ and the dispersion is quartic, $\tilde\Delta-\tilde\Delta_0\sim\gamma(k-k_0)^4$, the onset is modified to $\Im\tilde\chi\sim(\Omega-\tilde\Delta_0)^{-3/4}$. As it was discussed above, the geometry of gap is changed for $F_\tau>F^{({\rm cr})}$ by forming in the gap shape two minima at $k=\tilde k_\pm$ with corresponding magnitude $\tilde\Delta_{10}(\tilde k_\pm,F_t)=\min_k\tilde\Delta_{10}$ and a local maximum at the high-symmetry point $k_0=\pi/R$ (i.e., the $X$ point). The appearance of these features in the geometry of the laser-dressed gap leads to changing absorption spectra profile. 

\begin{figure*}[ht]
\centering
\includegraphics[width=0.99\linewidth]{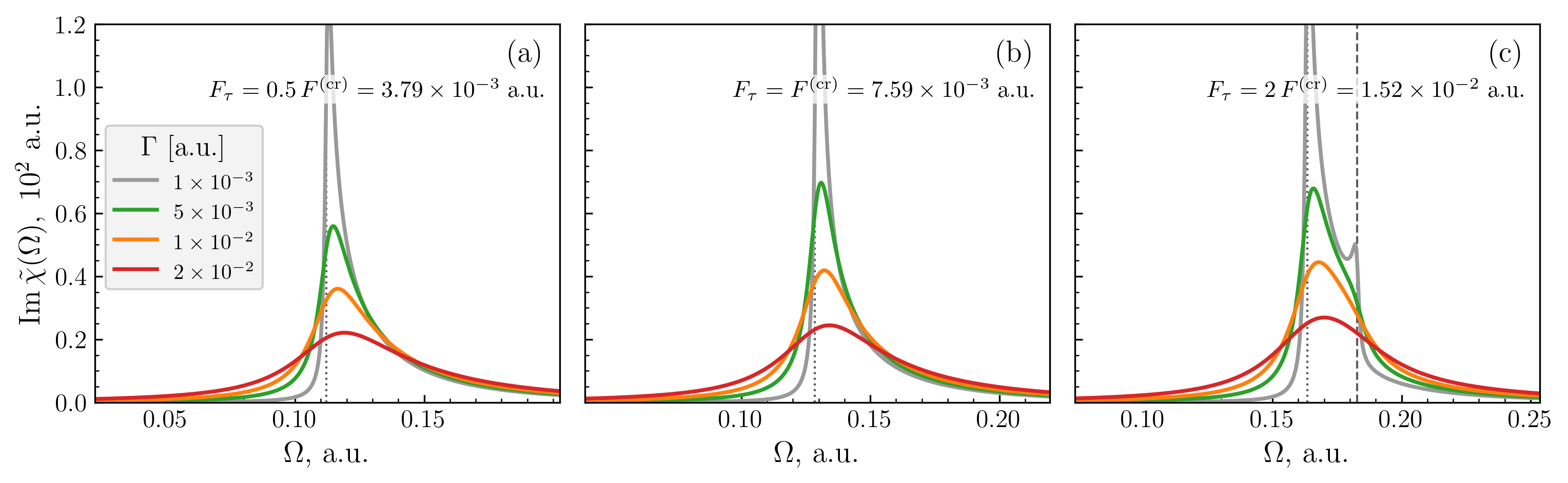}
\caption{Imaginary part of the instantaneous interband susceptibility $\tilde\chi(\Omega;F_\tau)$ [Eq.~\eqref{chi-inst}] for three instantaneous field strengths in the Kronig--Penney model; different curves correspond to the indicated phenomenological broadening $\Gamma$. Dotted lines mark the absorption edge $\tilde\Omega=\min_k\tilde\Delta_{10}(k,F_\tau)$; in panel (c) the dashed line marks $\Omega_X=\tilde\Delta_{10}(k_0,F_\tau)$, the dressed-gap value at the field-induced local maximum at $X$. Below $F^{(\rm cr)}$ a single edge feature is present; at $F^{(\rm cr)}$ the quartic critical dispersion modifies the spectral onset; above $F^{(\rm cr)}$ an additional feature emerges at $\Omega_X$, providing a spectroscopic signature of the gap bifurcation. Model parameters as in Fig.~\ref{fig:bandgap}.}
\label{fig:chi}
\end{figure*}

We illustrate in Fig.~\ref{fig:chi} the evolution of the spectral response across the bifurcation. For $F_\tau=0.5F^{(\rm cr)}$, the response exhibits a single absorption edge at $\tilde\Omega=\tilde\Delta_{10}(k_0,F_\tau)$, consistent with the presence of a single gap minimum. At the critical field $F_\tau=F^{(\rm cr)}$, the spectral profile changes qualitatively as a consequence of the flattening of the gap at the critical extremum. For the field above the critical value, $F_\tau=2F^{(\rm cr)}$, in addition to the edge at $\tilde\Omega=\min_k\tilde\Delta_{10}(k,F_\tau)$, the response exhibits a second feature at $\Omega_X=\tilde\Delta_{10}(k_0,F_\tau)$, whose position coincides with the dressed-gap value at the field-induced maximum at the $X$ point. The appearance of this feature, together with the edge associated with the two newly formed minima, directly reflects the restructured extremal structure of the dressed gap. The secondary feature remains clearly resolved in the present model for broadenings $\Gamma\lesssim10^{-2}$~a.u.

Our analysis within the Kronig--Penney model reveals two distinct regimes of the bifurcation. In narrow-gap materials with weak interband coupling, the critical field is low, but the bifurcation is confined to a narrow region of the Brillouin zone, making its spectroscopic signature relatively weak. In contrast, wide-gap materials with strong interband coupling (large potential amplitude) require higher critical fields, but the bifurcation spans a broader region of momentum space, resulting in a redistribution of optical weight and a stronger van Hove feature in the interband susceptibility. This suggests that for experimental observation via transient optical spectroscopy, materials with a strong interband mixing may be preferable, even though they require higher fields.

We emphasize the scope of this analysis: Eq.~\eqref{chi-inst} is an instantaneous adiabatic probe model and does not represent the full nonequilibrium pump--probe response; the power-law singularities discussed above are specific to the one-dimensional model. In higher dimensions they are softened---in three dimensions, the divergences are replaced by step-like or weaker features---but the emergence of an additional spectral feature at $\Omega_X$ and its energetic position, determined by the dressed-gap geometry, are dimension-independent. Thus, the number and energetic positions of the dominant spectral features provide a direct probe of the extremal structure of the instantaneous dressed gap.


\section{Conclusions}

In this work, we have developed an analytical framework for the laser-dressed electronic structure of solids in the intense low-frequency regime. By constructing an instantaneous adiabatic basis that incorporates nonperturbative interband mixing into the electronic structure itself, we derived a local normal form for the dressed interband gap, in which its local geometry is encoded by quadratic and quartic gap tensors $\tens{\upbeta}_{10}$ and $\tens{\upgamma}_{10}$. The central result is that strong-field dressing can change not only the magnitude of an interband gap, but also the number and character of its local extrema. At a critical field ($F^{(\rm cr)}$), the quadratic gap tensor changes its relevant eigenvalue structure, causing a single symmetric gap extremum to bifurcate into two symmetry-related extrema, while the original extremum changes its character. We identify the microscopic origin of this transition in the field-induced modification of the dressed-band curvatures and the redistribution of interband polarizations among multiple bands. Thus, the bifurcation is a genuinely multiband dressing effect rather than a uniform Stark renormalization of the gap.

The normal form is dictated by the symmetry constraints on the local gap geometry, while the bifurcation occurs when the field drives the relevant component of the quadratic gap tensor through the critical condition. It therefore represents a generic bifurcation mechanism available in inversion-symmetric periodic systems, rather than a feature specific to the Kronig–Penney model used for numerical illustration. Numerical analysis confirms this picture, showing that the bifurcation occurs across multiple high-symmetry points and band pairs.

We further demonstrated that this instantaneous laser-induced modification of the band geometry can provide a spectroscopic signature. It produces a distinct spectroscopic fingerprint in the form of an additional van Hove feature in the instantaneous interband susceptibility, whose position $\Omega_X$ directly encodes the value of the dressed gap at the field-induced local maximum. In this way, the extremal structure of the instantaneous gap becomes spectroscopically accessible without reconstructing the full band dispersion. 

Since our formalism relies solely on discrete translational symmetry, these results apply broadly to any spatially periodic system driven by strong low-frequency fields, including not only crystalline solids but also, for example, photonic lattices and cold atoms in optical lattices. More generally, the instantaneous spectral picture and the field-controlled gap geometry established here provide a natural starting point for a nonadiabatic theory of interband transitions in the strong low-frequency regime. This opens new avenues for understanding high-harmonic generation, strong-field spectroscopy, and attosecond dynamics in driven quantum materials.

\acknowledgments

This work is supported by the Chinese Academy of Sciences (YSBR-115), the National Natural Science Foundation of China (12374322, 12234002, and 12474486), CAS President's International Fellowship Initiative No. 2026PG0002 and 2026PVB0128, Russian Science Foundation Grant No. 26-11-00370 (in part of the developing theory of strong field interaction with a solid). 

\appendix
\section{\label{app:model} Analytical expressions for the Kronig-Penney model}

We illustrate our adiabatic theory of the laser-dressed electronic structure in solids using the simplest 1D Kronig--Penney model. Note that this model has been previously employed to study high-harmonic generation mechanisms in solids~\cite{NavarretePRA19}. Here, we apply it within the developed adiabatic formalism to account for instantaneous interband couplings in the laser-dressed states.

The crystal potential is given by the Dirac comb with period $R$ and strength $g$:
\begin{equation}
\label{KPM:U}
U(x)= g \sum_{l=-\infty}^{\infty}\delta(x-lR).
\end{equation}
For the one-dimensional potential defined in Eq.~\eqref{KPM:U}, all vectors and tensors become scalars.

The Kronig-Penney equation for the dispersion law $E_n(k)$ (for the $n$-th band) is
\begin{equation}
\label{KPM:disp}
\cos kR = \cos p_nR + \frac{g}{p_n}\sin{p_nR},\quad p_n=\sqrt{2E_n}.
\end{equation}

For the potential~\eqref{KPM:U}, the periodic Bloch solutions $u_{k,n}(x)$ [$u_{k,n}(x+R)=u_{k,n}(x)$, $0\leqslant x\leqslant R$] of the stationary (laser-field-free) Schr\"{o}dinger equation can be presented in the form:    
\begin{subequations}
\begin{multline}
\label{KPM:bloch-state}
u_{k,n}(x)= C_ne^{-ikx}\big[e^{ikR}\sin (p_nx) \\
- \sin \big(p_n(x-R)\big)\big],
\end{multline}
\begin{equation}
\label{KPM:norm}
C_n=\frac{(-1)^{\left\lfloor(p_n+k)\frac{R}{2\pi}\right\rfloor}}{\sqrt{\left( 1+\frac{g}{p_n^2R} \right)\sin^2 (p_nR) - \frac{g}{2p_n}\sin (2p_nR)}},
\end{equation}
\end{subequations}
where the phase of the normalization factor $C_n$ in Eq.~\eqref{KPM:norm} ensures the correct behavior of the Bloch state near the band edges at $(k+p_n)R=\pi l$ ($l$ is an integer), and $\lfloor x\rfloor$ is the greatest integer less than or equal to $x$.

With the analytical expression~\eqref{KPM:bloch-state}, we can find the explicit form for the field-free dipole matrix element $d_{nn'}(k)$, defined in Eq.~\eqref{dnn'}. Due to the inversion symmetry of the Kronig-Penney model, the diagonal elements $d_{nn}(k)$ are equal to zero, while for $n\neq n'$ we have:
\begin{equation}
\label{KPM:dnn'}
d_{nn'}(k) = iC_nC_{n'}\frac{p_{n}p_{n'}(\cos p_{n}R - \cos p_{n'}R)\sin kR }{R\Delta_{nn'}^2(k)},
\end{equation}
where $\Delta_{nn'}(k)=E_n(k)-E_{n'}(k)$. Note that for the considered model, the Berry connection $V_{nn}(t)=F(t)d_{nn}(K_t)$ disappears, while the instantaneous interband coupling $\hat{V}(t)$ for laser-dressed bands $\cE_n(t)$ can be taken into account in higher orders of the perturbation theory [cf. Eq.~\eqref{D:dressed:1}].

For inversion-symmetric bands over the region $-\pi<kR\leqslant \pi$ (as in the Kronig--Penney model), the local minimum $k=\tilde{k}$ of the gap $\Delta_{10}(k)$ between the valence band and the nearest conduction band is at $\tilde{k} = \pi/R$.

\end{document}